# Enabling Real-Time Volumetric Imaging in Interventional Radiology Suits via a Deep Learning Framework Robust to C-arm Tilt

Fawazilla Utomo, Tess Reynolds, Nicholas Hindley

The University of Sydney

## ABSTRACT

Contemporary interventional imaging lacks the real-time 3D guidance needed for the precise localization of mobile thoracic targets. While Cone-Beam CT (CBCT) provides 3D data, it is often too slow for dynamic motion tracking. Deep learning frameworks that reconstruct 3D volumes from sparse 2D projections offer a promising solution, but their performance under the geometrically complex, non-zero tilt acquisitions common in interventional radiology is unknown. This study evaluates the robustness of a patient-specific deep learning framework, designed to estimate 3D motion, to a range of C-arm cranial-caudal tilts. Using a 4D digital phantom with a simulated respiratory cycle, 2D X-ray projections were simulated at five cranial-caudal tilt angles (0°, ± 15°, ± 30°) across 10 breathing phases. A separate deep learning model was trained for each tilt condition to reconstruct 3D volumetric images. The framework demonstrated consistently high-fidelity reconstruction across all tilts, with a mean Structural Similarity Index (SSIM) > 0.980. While statistical analysis revealed significant differences in performance between tilt groups ($p < 0.0001$), the absolute magnitude of these differences was minimal (e.g., the mean absolute difference in SSIM across all tilt conditions was ~0.0005), indicating they were not functionally significant. The magnitude of respiratory motion was found to be the dominant factor influencing accuracy, with the impact of C-arm tilt being a much smaller, secondary effect. These findings demonstrate that a patient-specific, motion-estimation-based deep learning approach is robust to geometric variations encountered in realistic clinical scenarios, representing a critical step towards enabling real-time 3D guidance in flexible interventional settings.

**Keywords:**

## 1. INTRODUCTION

Accurate, real-time 3D image guidance offers crucial depth information that is absent in contemporary interventional imaging. In fields like interventional radiology and radiation therapy, achieving precise 3D localization of mobile thoracic targets is of utmost importance. While conventional 3D modalities like Cone-Beam CT (CBCT) can provide this depth information, they have several disadvantages which limit their utility for real-time image guidance: a speed that is too slow for dynamic tracking of respiratory motion, susceptibility to significant motion artifacting, and a high radiation dose associated with repeated acquisitions.

To overcome this, several patient-specific, deep learning frameworks have been developed to reconstruct full 3D volumetric images from a sparse set of 2D projections, addressing the problem of tracking respiratory motion. One such example of this feasibility was demonstrated by Shen et al., who showed that a patient-specific model could learn the mapping from a single 2D projection to a 3D CT volume [2]. Other works have explored the usage of other various deep learning architectures, including Generative Adversarial Networks (GANs) to improve realism [3] and direct-to-volume Convolutional Neural Networks (CNNs) [4]. However, the translation of these techniques to broader clinical application is hindered by a significant knowledge gap: nearly all existing 2D-to-3D reconstruction models have been developed under the highly controlled conditions of radiotherapy, which almost exclusively utilize standard, zero-tilt image acquisitions. This geometric constraint does not reflect the reality of modern interventional radiology, where robotic C-arms are used at non-zero cranial-caudal tilt angles to obtain optimal, unobstructed views of the target anatomy. It is currently unknown whether the high accuracy reported for these deep learning models can be maintained when tasked with real-time volumetric imaging at arbitrary gantry angles.

This study was designed to address this gap by providing the first evaluation of a patient-specific deep learning model across a range of cranial-caudal C-arm tilts. It was hypothesized that Voxelmap, a deep-learning model designed to predict the underlying 3D deformation vector field (DVF) of 2D images [5], would generate 3D volumetric images from moderate tilted 2D images (up to ±30°) with no functionally significant degradation in image quality compared to a zero-tilt baseline, provided that essentially anatomy, such as the diaphragm, is left unobstructed.

## 2. METHODS

### 2.1 Data Source and Simulation

The study was performed using the 4D Extended Cardiac-Torso (XCAT), which provides a ground-truth 3D anatomical model [6]. This ground-truth was then divided into ten distinct volumetric datasets corresponding to ten phases of a simulated respiratory cycle. From this sequence, the 3D XCAT volume at peak exhalation (Phase 6) was designated as the reference volume. The ground-truth motion between respiratory states was defined by 3D DVFs, which were calculated by registering the nine other motion-state volumes to this reference volume using the Elastix toolkit [7].

To create the experimental dataset for this study, 2D X-ray projections were simulated at five distinct cranial-caudal tilt angles: -30°, -15°, 0°, +15°, and +30°. For each tilt condition, the 3D XCAT volumes for all respiratory phases were rotated in silico around the anterior-posterior axis. Subsequently, for each of the ten respiratory phases and five tilt angles a full 360° gantry rotation was simulating using a forward projection model. This process yielded a total dataset of 31,500 unique 2D projection images, creating five experimental groups with a sample size of n=6,300 projections per group.

### 2.2 Voxelmap Framework and Training

A patient-specific deep learning framework (Voxelmap) designed to predict a 3D DVF from a pair of co-planar 2D projections was then employed. The network utilizes a CNN-based encoder-decoder architecture to learn the non-linear mapping between a "reference" 2D projection (from the peak-exhale phase) and an "acquired" 2D projections (from any other respiratory phase) to the ground-truth 3D DVF that describes the motion between those two states. The final 3D volumetric image is generated by warping the 3D reference volume with the network's predicted DVF.

### 2.3 Data Pre-processing and Training

Prior to training, all data was downsampled due to GPU memory constraints. The 3D CT volumes were resized to a resolution of 128x128x128 voxels and all 2D projections were downsampled to 128x128 pixels.

For each of the five tilt conditions, a separate, patient-and-angle-specific model was trained. The training process utilized the simulated 2D projections from 9 of the 10 respiratory phases, with the remaining phase held out for testing. This cross-validation was repeated 10 times to ensure each phase as used for testing once, allowing for a robust evaluation of the model's performance.

### 2.4 Performance Evaluation and Statistical Analysis

The primary outcomes were four quantitative metrics of volumetric imaging accuracy: Mean Absolute Error (MAE), Root-Mean-Squared Error (RMSE), Structural Similarity Index Measure (SSIM), and Peak Signal-to-Noise Ratio (PSNR). These metrics were calculated by comparing the network's predicted 3D volume to the known ground-truth 3D volume for the corresponding respiratory phase.

To determine if C-arm tilt significantly impacted performance, a one-way analysis of variance (ANOVA) was conducted for each metric to compare mean performance across the five tilt groups within each breathing phase. A p-value < 0.05 was considered statistically significant. For significant results, a Tukey's HSD post-hoc test was performed to identify which specific tilt angle pairs differ.

## 3. RESULTS

### 3.1 Influence of Respiratory Motion and Gantry Angle

The Voxelmap framework demonstrated consistently high-fidelity 3D image reconstruction across all five tested tilt angles. The overall quantitative performance in summarized in Table 1. The mean SSIM remained excellent (>0.980), and error metrics were consistently low (mean MAE ~ 0.016) for all tilt groups, signifying a high degree of accuracy between the predicted and ground-truth volumes.

The distribution of these performance metrics for each tilt angle is visualized in Figure 1. The one-way ANOVA confirmed a statistically significant effect of tilt angle on all four metrics (p < 0.0001). However, the absolute magnitude of these differences was minimal, suggesting a lack of functional significance. For example, in the phase of maximum respiratory motion (Phase 1), the largest observed mean difference in SSIM between any tilt groups was only 0.0024, and the largest mean difference in MAE was 0.0015. This suggests that while the model's performance does vary measurability with tilt, the changes are not functionally significant from an imaging quality perspective.

Table 1. Overall Performance of Voxelmap XCAT Reconstruction for Different C-Arm Tilt Angles. Values are presented as Mean ± Standard Deviation

| Tilt (°) | n | MAE | RMSE | SSIM | PSNR (dB) |
|---|---|---|---|---|---|
| -30 | 6300 | 0.0164 ± 0.0062 | 0.0590 ± 0.0191 | 0.9812 ± 0.0095 | 34.32 ± 4.69 |
| -15 | 6300 | 0.0158 ± 0.0065 | 0.0592 ± 0.0200 | 0.9811 ± 0.0095 | 34.69 ± 6.16 |
| 0 | 6300 | 0.0163 ± 0.0068 | 0.0603 ± 0.0210 | 0.9807 ± 0.0099 | 34.45 ± 5.72 |
| 15 | 6300 | 0.0166 ± 0.0063 | 0.0611 ± 0.0197 | 0.9805 ± 0.0100 | 33.93 ± 4.36 |
| 30 | 6300 | 0.0158 ±0.0065 | 0.0583 ± 0.0200 | 0.9814 ± 0.0095 | 34.75 ± 5.82 |

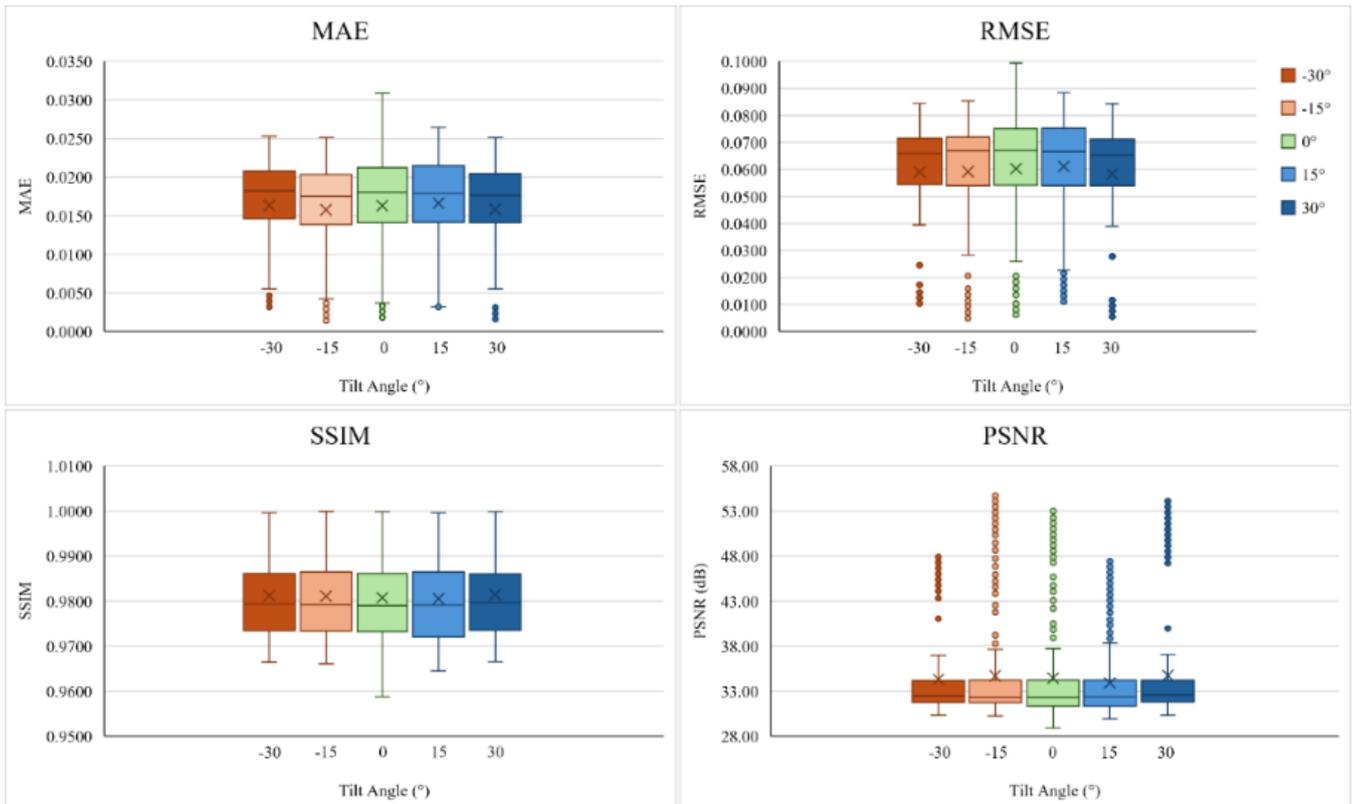

Figure 1. Volumetric Imaging Accuracy of Voxelmap XCAT Reconstruction Across Different C-Arm Tilt Angles. Distribution of four key accuracy metrics across five cranial-caudal tilt angles.

### 3.2 Influence of Respiratory Motion and Gantry Angle

Analysis of performance across the respiratory cycle revealed that the magnitude of respiratory motion was the dominant factor influencing accuracy. As shown in Figure 2, imaging accuracy for all tilt angles peaked at the zero-motion reference state of peak exhale (Phase 6), achieving near-perfect mean SSIM (≈ 0.999), and degraded predictably with increasing motion displacement. Crucially, the performance curves for all five tilt groups remained tightly clustered, visually confirming that the impact of C-arm tilt is minimal compared to the effect of respiratory motion.

Analysis of performance by gantry angle revealed that reconstruction accuracy varied, with the poorest mean performance consistently observed in the lateral projection bin centred at 270°. This angle corresponds to a view with minimal anterior-posterior structural information to guide the model.

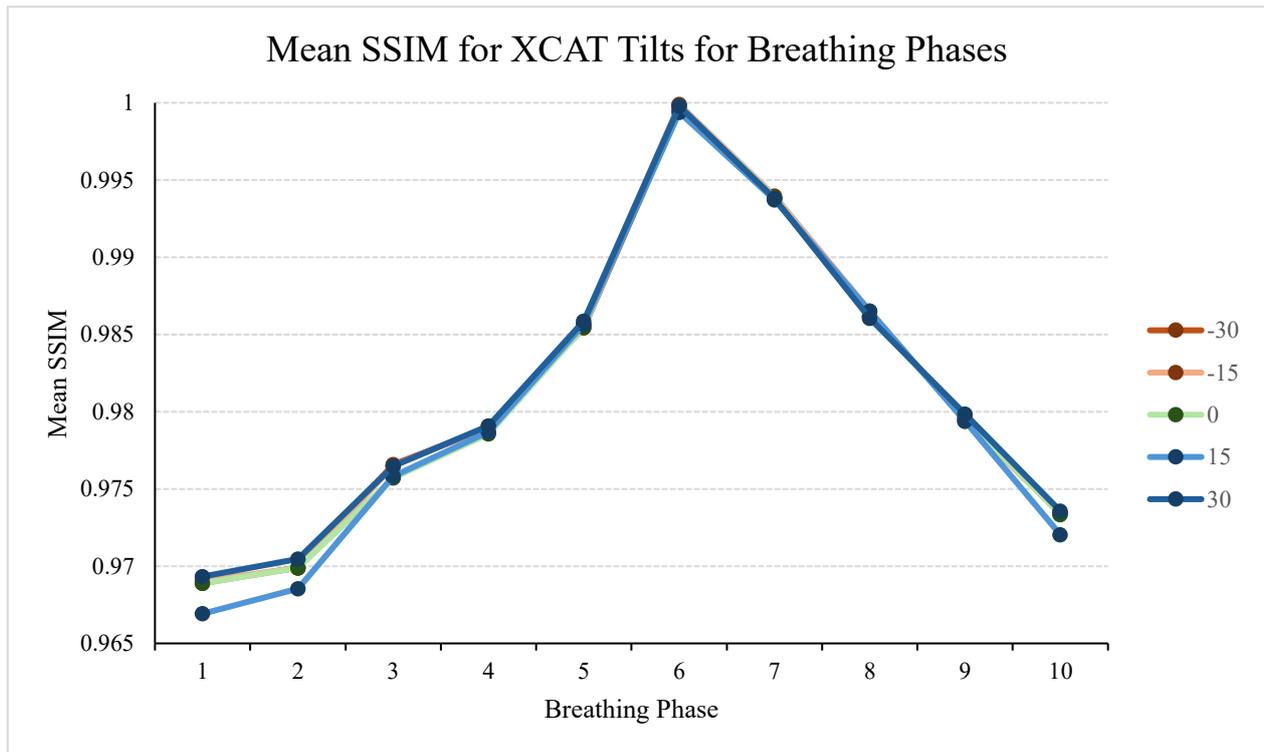

Figure 2. Influence of Respiratory Motion and C-Arm Tilt on Reconstruction Accuracy. Median Structural Similarity Index Measure (SSIM) is plotted against the ten respiratory phases for each of the five cranial-caudal tilt angles.

### 3.3 Qualitative Performance

A qualitative assessment is shown in Figure 3. Despite the clear geometric variations in the input 2D projections caused by the cranial-caudal tilt, visual inspection of the predicted 3D volumes reveals minimal degradation in image quality, with predicted coronal slices being indistinguishable from their corresponding ground-truth slices. The absolute difference maps confirms that reconstruction errors are minimal and of low magnitude.

| Tilt (°) | Input 2D Projection (at 270° Gantry Angle) | Ground-Truth 3D Volume (Coronal Slice) | Voxelmap Predicted 3D Volume (Coronal Slice) | Absolute Difference Map |
|---|---|---|---|---|

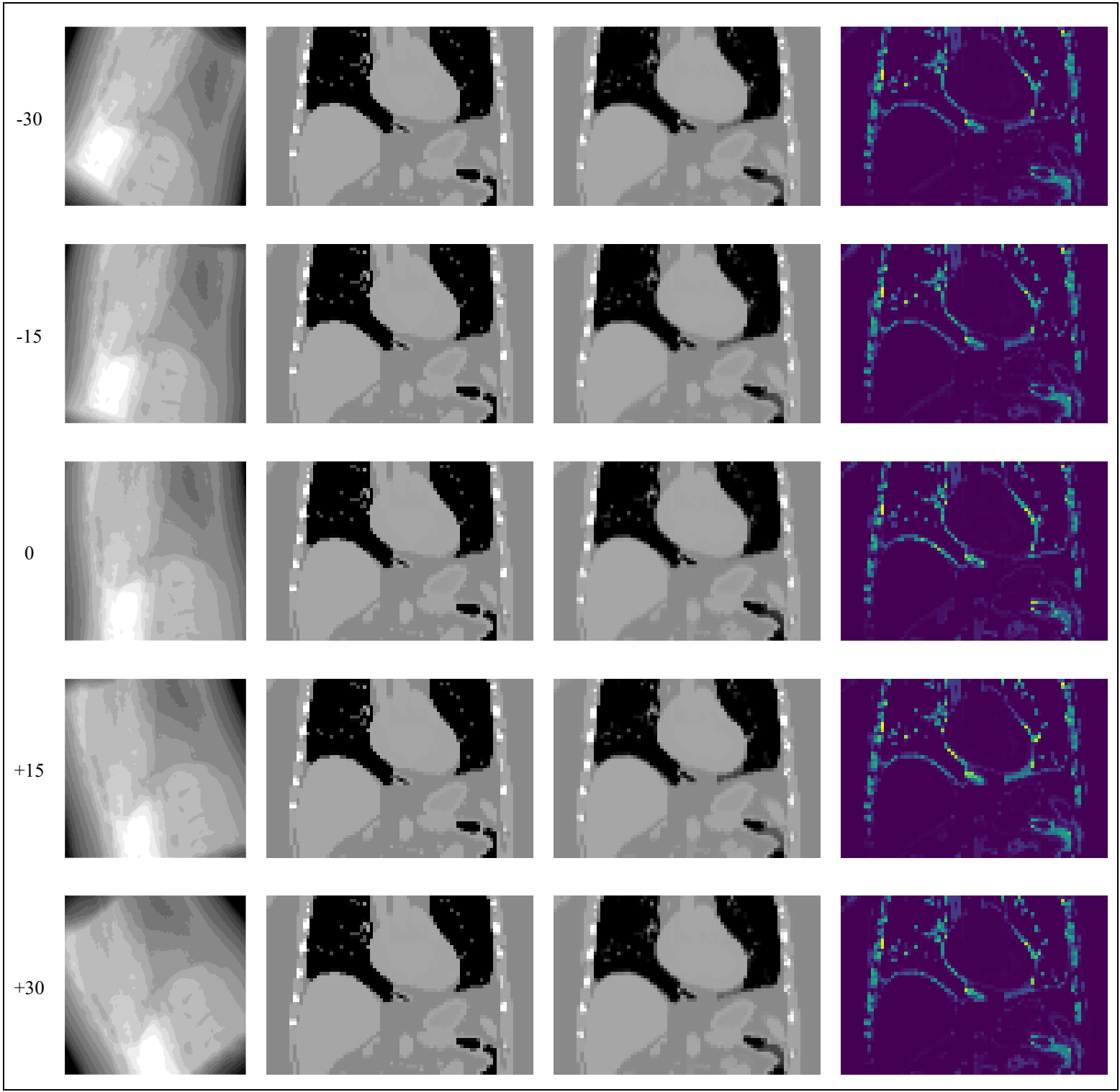

Figure 3. Qualitative Comparison of Voxelmap Reconstructions from Geometrically Varied 2D Projections. This figure demonstrates the high fidelity of the Voxelmap framework for a representative case (Phase 1, 270° gantry angle) across five C-arm tilt conditions.

## 4. DISCUSSION AND CONCLUSION

This study found that a patient-specific deep learning framework can generate a high-fidelity 3D volumetric reconstruction from 2D projections acquired with C-arm cranial-caudal tilts up to ± 30°. This principal finding was robust across all tested respiratory phases and was supported by high performance across four key imaging metrics. This result directly addresses a knowledge gap, as previous work in this domain has been almost exclusively in acquisitions with zero cranial-caudal tilt [2, 4]. The distinction between statistical significance and practical effect size is a key outcome; while our large dataset allowed for the detection of minute, statistically significant differences between tilt groups, the absolute magnitude of these variations was negligible, confirming our hypothesis that Voxelmap is functionally robust to moderate tilts.

The robustness of the framework can be interpreted through the lens of its underlying architecture. Where other models have focused on direct 3D volume generation [2, 3, 4], Voxelmap's approach of first predicting a 3D DVF appears robust to geometric variance. Effectively, this means the network learns to disentangle the true 3D anatomical deformation caused by respiration from the apparent 2D distortion caused by the tilted C-arm views. This is a significant achievement, as the 2D projections from different tilts can appear quite different, yet the model consistently predicted the correct 3D motion field.

The analysis across the respiratory cycle further demonstrated that the phase of the breathing motion was the dominant factor influencing accuracy, with the impact of tilt being a much smaller, secondary effect. This supports the hypothesis that if key anatomical drivers of motion, such as the diaphragm, remain visible in the 2D projection, the model can reliably infer the correct 3D motion.

While the findings from this study suggest that real-time 3D guidance from a flexibly positioned C-arm is a feasible goal using existing hardware, the study's in-silico design presents several limitations. These include the use of a single, idealized phantom with noise-free projections and periodic, regular respiratory motion, which do not reflect the complexity of real-world clinical scenarios. Therefore, crucial future research must focus on validating these findings in more realistic settings. This includes assessing performance with physical phantoms to account for scatter and noise, and training the framework on cohorts of retrospective patient 4D-CT scans to ensure robustness against anatomical diversity and irregular breathing patterns.

In conclusion, this study demonstrates that a patient-specific deep learning framework can be robust to significant variation in C-arm imaging geometry. By showing that generation of high-fidelity 3D volumetric reconstruction is possible from tilted 2D projections, this study bridges the gap between the original use case of these models in radiotherapy imaging and the flexible needs of the interventional radiology suite.

## ACKNOWLEDGEMENTS

This research was done as a project under the University of Sydney's Year 3 Medical Doctor program, under the supervision of Dr. Nicholas Hindley and Dr. Tess Reynolds.

## REFERENCES

[1] Katano A., Minamitani M., Ohira S., and Yamashita H., "Recent Advances and Challenges in Stereotactic Body Radiotherapy," Technol. Cancer Res. Treat. **23,** 15330338241229363 (2024). https://doi.org/10.1177/15330338241229363

[2] Shen, L., Zhao, W. and Xing, L., "Patient-specific reconstruction of volumetric computed tomography images from a single projection view via deep learning," Nat. Biomed. Eng. **3**(11), 880–888 (2019). https://doi.org/10.1038/s41551-019-0466-4

[3] Lei Y., Tian Z., Wang T., Higgins K., Bradley J.D., Curran W.J., Liu T., and Yang X., "Deep learning-based real-time volumetric imaging for lung stereotactic body radiation therapy: a proof of concept study," Phys Med Biol **65**(23), 235003 (2020). https://doi.org/10.1088/1361-6560/abc303

[4] Loyen E., Dasnoy-Sumell D., and Macq B., "3DCT reconstruction from single X-ray projection using Deep Learning," Radiother. Oncol. **170**(Supplement 1), S171 EP-S172 (2022). https://doi.org/10.1016/S0167-8140(22)02317-9

[5] Hindley N., Shieh C.-C., and Keall P., "A patient-specific deep learning framework for 3D motion estimation and volumetric imaging during lung cancer radiotherapy," Phys Med Biol **68**(14) (2023). https://doi.org/10.1088/1361-6560/ace1d0

[6] Segars, W. P., Bond, J., Frush, J., Hon, S., Eckersley, C., Williams, C. H., Feng, J., Tward, D. J., Ratnanather, J. T., Miller, M. I., Frush, D. and Samei, E., "Population of anatomically variable 4D XCAT adult phantoms for imaging research and optimization," Medical Physics **40**(4), 043701 (2013). https://doi.org/10.1118/1.4794178

[7] Klein, S., Staring, M., Murphy, K., Viergever, M. A. and Pluim, J. P. W., "elastix: A Toolbox for Intensity-Based Medical Image Registration," IEEE Transactions on Medical Imaging **29**(1), 196–205 (2010). https://doi.org/10.1109/tmi.2009.2035616